\documentstyle{article}
  \begin{document}
  \title{Checking the variability of the gravitational constant with binary pulsars}
  \author{G.S. Bisnovatyi-Kogan
  \thanks{Institute of Space Research (IKI), Moscow, Russia, Profsoyuznaya
  84/32, Moscow 117810, Russia,  gkogan@mx.iki.rssi.ru}}
  \date{}
  \maketitle
\begin{abstract}
The most precise measurements are done at present by timing of
radiopulsars in binary systems with two neutron stars. The timing
measurements of the Taylor-Hulse pulsar B1913+16 gave the most precise
results on testing of general relativity (GR), finding implicit
proof of existence of gravitational waves. We show that available results of existing
measurements, obtained to the year 1993,
in combination with the results of the Mariner 10 in (1992),
give the boundaries for the variation
of the gravitational constant ${\dot G}/{G}$ inside the limits $(-0.6 \div +2)\cdot
10^{-12}$ year{$^{-1}$}.
\end{abstract}

\smallskip


\smallskip

Since 1937 \cite{miln37},\cite{dir37}, and up to present times \cite{uzan},\cite{fri04}
the possibility of variation of fundamental physical constants arise interest among physicists.
While theories exist where these constants may vary \cite{wu86},\cite{var95},\cite{meln1},
\cite{meln2}, it is
evident that only from experiments, or better to say from the astrophysical observations,
we may obtain a realistic answer. The spectral observations of distant quasars have been used
for estimation of upper limits for the variations of the fine structure constant,
electron and proton masses \cite{var95}.

Variations of the gravitational constant $G$ have been measured by different methods
\cite{uzan}, among which are
investigations of stellar and planetary orbits.
The measurements on the Viking Lander ranging data gave the limits
\cite{hel83} for the variation of the gravitational constant in the form

\begin{equation}
\label{viking}
{\dot G}/{G}=(2\pm 4)\cdot 10^{-12} {\rm yr}^{-1},
\end{equation}
and the combination of Mariner 10 and Mercury and Venus ranging gave \cite{and92}

\begin{equation}
\label{mariner}
{\dot G}/{G}=(0\pm 2)\cdot 10^{-12} {\rm yr}^{-1},
\end{equation}
Lunar Laser Ranging experiments are lasting for many years, and improve the estimations of
${\dot G}/{G}$ from ${\dot G}/{G}< 6\cdot 10^{-12} {\rm yr}^{-1}$
in \cite{dic94} the year 1994, up to

\begin{equation}
\label{lunar}
{\dot G}/{G}=(0.4\pm 0.9)\cdot 10^{-12} {\rm yr}^{-1},
\end{equation}
in \cite{will04} the year 2004.

Recycled pulsars give now the best available timing precision. Especially important
results may be obtained from timing measurements of the pulsars in close binaries
consisting of two neutron stars. These measurements had been already used for
checking general relativity \cite{tay93} (PSR B1913+16),
and \cite{lyne04},\cite{kra05} (PSR J0737-3039). From the long time
observations of the the pulsar PSR B1913+16 changes in the binary period
had been measured, and explaining these changes by emission of gravitational waves,
it was found \cite{tay94}, that
"Einstein's theory passes this extraordinary stringent test  with a fractional
accuracy better than 0.4\%". Similar accuracy was obtained from only one year observations
of the close binary consisting of two pulsars PSR J0737-3039A and B \cite{lyne04}
\cite{kra05}. Farther
observations of this system should considerably improve the precision, in comparison with
the one reached by PSR B1913+16.

The restrictions on the gravitational theory, and checking of its validity is possible
because  timing permits to measure several quantities what gives an overlapping
information. With the two masses as the only free parameters, the measurement of
three or more post-keplerian parameters over-constrains the system, and thereby
provides a test ground for theories of gravity.  In a theory that
describes a binary system correctly, the post-keplerian parameters produce
theory-dependent lines in a mass-mass diagram that all intersect in a
single point \cite{dd86},\cite{kra05}.
Measurements of 5 relativistic parameters together with independent
determination of the mass ratio, due to presence
of the second pulsar, are available in PSR J0737-3039, in comparison with only 3 parameters in
the PSR B1913+16.

The variation of fundamental constant could be incorporated in the general procedure
of the timing data development. With 6 measurable quantities at present
(and possibility to increase this number in future \cite{kra05}) variability of several
fundamental constant can be investigated.

Nevertheless, even without any complicated calculations  strong restrictions can be obtained
to the variation of the gravitational constant using only the measurements of the variations
of the binary period. Change of $G$ has the most evident influence on the
orbital motion, because only emission of the gravitational waves compete here with the variation of
$G$. Influence of $G$ variation of the pulsar period changes, star contraction at increasing
gravity, or star expansion at its decreasing, accompanying by corresponding period changes at constant
angular momentum, are strongly contaminated with the intrinsic changes of the rotational
period, due to losses of the rotational energy and angular momentum.
A theoretical description of these losses
is rather poor \cite{mic00}.
If we accept the correctness of general relativity inside the observed accuracy,
than we can use the residuals (error box) of the measurements of
decay of the binary period $\dot P_b$ for the estimation of variations of $G$.
Such estimations have been done first in \cite{dam88}, and later in \cite{kaspi},
using timing data from \cite{tay93} with the result

\begin{equation}
\label{kasp}
{\dot G}/{G}=(4\pm 5)\cdot 10^{-12} {\rm yr}^{-1}.
\end{equation}
Here we make slightly more careful analysis of the data in \cite{tay93}, and combine it
with Mariner 10 data, to decrease the error box for
${\dot G}/{G}$, in comparison with \cite{kaspi}.
Influence of gravitational waves, and $G$ variations on the binary period may be
considered separately in linear approximation due to smallness of both effects.
The orbital angular momentum is conserved during slow $G$ variations, as an adiabatic
invariant, so only total energy of the system is changing due to change of the
gravitational constant. In newtonian approximation the total energy $E$ and orbital
angular momentum $M$ of the binary in the mass center system are written as

\begin{equation}
\label{eq1}
(a)\,\,\,E=\frac{m\,v_{\phi a}^2}{2}-\frac{Gm_1 m_2}{a}, \quad (b)\,\,\,M=mav_{\phi a}.
\end{equation}
Here $m_1$, $m_2$ are masses, considered as points, $m=\frac{m_1 m_2}{m_1+m_2}$,
$v_{\phi a}$ is the azimuthal  velocity at apogee, where radial velocity is zero.
The parameters of the binary orbit, such as
large semi-axis $a$, eccentricity $e$, and orbital period $P_b$,
are uniquely determined by the conserved values $E<0$ and $M$ as
\cite{ll90}

\begin{equation}
\label{eq2}
(c)\,\,\,a=\frac{Gm_1 m_2}{2|E|},\quad (d)\,\,\,P_b=\pi Gm_1 m_2\sqrt{\frac{M}{2|E|^3}},\quad
(f)\,\,\, e=\sqrt{1-\frac{2|E| M^2}{G^2 m m_1^2 m_2^2}}.
\end{equation}
Slow variation of $G$ imply the variation of $P_p$, $a$, $v_{\phi a}$, $E$ and $e$,
which are connected by relations following from variations of corresponding relations in (\ref{eq1}),
\ref{eq2})

$$
(b)\,\,\,\frac{\delta v_{\phi a}}{v_{\phi a}}=-\frac{\delta a}{a},\,\,
(c)\,\,\,\frac{\delta |E|}{|E|}=\frac{\delta G}{G}-\frac{\delta a}{a},\,\,
$$
\begin{equation}
\label{eq3}
(a)\,\,\,\delta |E|=\left(m\,v_{\phi a}^2-\frac{Gm_1 m_2}{a}\right)\frac{\delta a}{a}+
\frac{Gm_1 m_2}{a}\frac{\delta G}{G};
\end{equation}

\begin{equation}
\label{eq4}
(d)\,\,\,\frac{\delta P_b}{P_b}=\frac{\delta G}{G}-\frac{3}{2}\frac{\delta |E|}{|E|},\quad
(f)\,\,\,\frac{\delta e}{e}=\frac{M^2}{G^2 m m_1^2 m_2^2}\frac{|E|}{e^2}
\left(\frac{\delta G}{G}-\frac{\delta |E|}{|E|}\right).
\end{equation}
It follows from (\ref{eq3})

\begin{equation}
\label{eq5}
\frac{\delta a}{a}=-\frac{\delta G}{G},\quad
\frac{\delta v_{\phi a}}{v_{\phi a}}=\frac{\delta G}{G},\quad
\frac{\delta |E|}{|E|}=2\frac{\delta G}{G},
\end{equation}
what, using (\ref{eq4}), gives

\begin{equation}
\label{eq6}
\frac{\delta P_b}{P_b}=-2\frac{\delta G}{G},\quad
\frac{\dot G}{G}=-\frac{1}{2}\frac{\dot P_b}{P_b},\quad
\frac{\delta e}{e}=0,\,\, e={\rm comst}.
\end{equation}
The first relation of (\ref{eq6}) was derived in \cite{dam88}
in a more complicated way.
So, from measurements of changes of the orbital period we may
estimate variations of the gravitational constant, after account of gravitational wave
reaction. For the pulsar PSR B1913+16 the
error budget for the orbital period derivative,
in comparison with the general relativistic prediction, is given in Table 1, from \cite{tay93}.

\begin{table}
 \caption{Error budget for the orbital period derivative,
in comparison with the general relativistic prediction, from \cite{tay93}.}
\medskip
\begin{tabular}{lll}
%\tableline
                       & Parameter  &($10^{-12}$)   \\
                       \hline
 Observed value        &  $\dot P_b^{obs}$  & -2.4225$\pm$0.0056   \\
 Galactic contribution & $\dot P_b^{gal}$   & -0.0124$\pm$0.0064  \\
 Intrinsic orbital period decay & $\dot P_b^{obs}$ - $\dot P_b^{gal}$&-2.4101$\pm$0.0085  \\
 General relativistic prediction & $\dot P_b^{GR}$ &  -2.4025$\pm$0.0001     \\
\hline \hline
\end{tabular}
\label{htpul}
\end{table}
It follows, that the observed values coincide with GR
prediction within the error bar
$$
(\dot P_b^{obs} - \dot P_b^{gal})/\dot P_b^{GR}\,=\, 1.0032\pm0.0035.
$$
Supposing that all deviations from GR are connected with $G$ variation,
we obtain the upper limit for these variation as follows

\begin{equation}
\label{bp}
\frac{\dot G}{G}=(4.3\pm 4.9)\cdot 10^{-12}\, {\rm yr}^{-1}.
\end{equation}
To obtain this result we have assumed correctness of GR prediction (last line in the
Table 1), used the difference between the observed (third line in Table 1) and theoretical
values ($-0.0076\cdot 10^{-12}$) for the result,
and the combined (theoretical + experimental) error ($\pm 0.0086 \cdot 10^{-12}$)
for the error box. To transform the nondimensional value of $\dot P$
into prediction for $\frac{\dot G}{G}$ in (\ref{bp}),
we should multiply the result and the error by $\frac{1}{2P_b}=565.4$ yr$^{-1}$,
where $P_b=0.323$ days = $8.84 \cdot 10^{-4}$ years. So, more accurate estimations
(\ref{bp}) slightly improve the values (\ref{kasp}), obtained in \cite{kaspi}.
This improvement narrows on 40\% the boundary in the negative region (most important from the
theoretical point of view \cite{wu86},\cite{meln1},\cite{meln2}), from $-1\cdot 10^{-12}$ to
$-0.6 \cdot 10^{-12}$.

Combination of two results (\ref{bp}) and (\ref{mariner}) permits
to narrow the range of $G$ variations, which is now may be situated in the limits

$${\dot G}/{G}\,\, {\rm is\,\, within \,\,the \,\,interval}
\,\,(-0.6, \,\,2)\cdot 10^{-12} {\rm yr}^{-1}.
$$
These limits are slightly shifted from predictions of superstring theories \cite{wu86}
${\dot G}/{G} \approx -1 \cdot 10^{-11\pm 1} {\rm yr}^{-1},
$
and permit for $G$ to remain constant. They are independent on the influence of tidal
interaction between Earth and Moon.
Regretfully, the data of measurements of $\dot P_b$ in PSR 1913+16 since 1993, as well as the
refined data on the PSR J0737-3039 are not available, and the
published \cite{kra05} errors in measurement of $\dot P_b$ in the last case are still
considerably larger than that
for PSR B1913+16 in 1993. Farther observations of both pulsars
could improve the precision in the estimation of $G$ variations.

\medskip

{\Large{\bf Acknowledgement}}

\medskip

Author is grateful to Dr. Marco Merafina and the Physical
Department of the University La Sapienca, for hospitality during
the work on this paper.

\end{document}